\documentclass[useAMS,usenatbib,usegraphicx,fleqn]{mn2e}
\usepackage{epsfig}
\usepackage{amsmath}
\usepackage{graphicx}
\usepackage{color}
\usepackage[normalem]{ulem}
\newcommand{\mnras}{MNRAS}
\newcommand{\aj}{AJ}
\newcommand{\apj}{ApJ}
\newcommand{\apjl}{ApJL}
\newcommand{\apjs}{ApJS}
\newcommand{\nat}{Nature}
\newcommand{\aap}{A\&A}

\newcommand{\araa}{ARAA}

\newcommand{\pasp}{PASP}
\newcommand{\planss}{PLANSS}

\title[Calcium-rich gap transients]{Calcium-Rich Gap Transients: Tidal Detonations of White Dwarfs?}
\author[Sell et al.]
{\parbox{\textwidth}{P.~H.~Sell$^1$\thanks{email: paul.sell@ttu.edu},
T. J. Maccarone$^1$,
R. Kotak$^2$,
C. Knigge$^3$,
D. J. Sand$^1$}\vspace{0.4cm}\\
$^1$Department of Physics, Texas Tech University, Lubbock TX 79409, USA \\
$^2$School of Physics and Astronomy, Queen's University Belfast \\
$^3$School of Physics and Astronomy, University of Southampton, Highfield, Southampton, SO17 1BJ, UK}

\begin{document}

\date{}

\pagerange{\pageref{firstpage}--\pageref{lastpage}} \pubyear{}

\maketitle

\label{firstpage}

\begin{abstract}

We hypothesize that at least some of the recently discovered class of
calcium-rich gap transients are tidal detonation events of white dwarfs (WDs) by
black holes (BHs) or possibly neutron stars.  We show that the properties of the
calcium-rich gap transients agree well with the predictions of the tidal
detonation model.  Under the predictions of this model, we use a follow-up X-ray
observation of one of these transients, SN~2012hn, to place weak upper limits on the
detonator mass of this system that include all intermediate-mass BHs (IMBHs).  As these
transients are preferentially in the stellar haloes of galaxies, we discuss the
possibility that these transients are tidal detonations of WDs caused by random flyby
encounters with IMBHs in dwarf galaxies or globular clusters.  This
possibility has been already suggested in the literature but without connection to
the calcium-rich gap transients.  In order for the random flyby cross-section to be
high enough, these events would have to be occurring inside these dense stellar
associations.  However, there is a lack of evidence for IMBHs in
these systems, and recent observations have ruled out all but the very faintest dwarf
galaxies and globular clusters for a few of these transients.  Another possibility
is that these are tidal detonations caused by three-body interactions, where a WD
is perturbed toward the detonator in isolated multiple star systems.  We highlight
a number of ways this could occur, even in lower-mass systems with stellar-mass BHs
or neutron stars.  Finally, we outline several new observational tests of this
scenario, which are feasible with current instrumentation.

\end{abstract}


\section{Introduction}
\label{section:intro}

The field of time domain astronomy has recently exploded with the detection of
many new classes of transient sources.  Considerable progess has so far been
made at optical wavelengths with the advent of a host of new programs designed
to search for sources at a variety of cadences.  They are also more sensitive to
finding sources at different luminosities in the gap between that of novae and
traditional supernovae \citep{kasliwal11}.  Some of the best-known current
transient surveys are at optical wavelengths:  the Palomar Transient Factory
\citep[PTF;][]{rau09}, the Panoramic Survey Telescope and Rapid Response
System \citep[Pan-STARRS;][]{hodapp04}, and the Public European Southern
Observatory Spectroscopic Survey for Transient Objects \citep[PESSTO;][]{smartt14}.
New observatories (e.g. the Large Synoptic Survey Telescope), including some at other
wavelengths (e.g. the Square Kilometer Array, eRosita), are being planned as well.
These observatories enable us to better map out the parameter space of known
transients as well as discover new classes of transients.

One of the recently discovered classes is ``calcium-rich gap
transients'' --- so called because their spectra contain strong forbidden and
permitted calcium lines, and they have peak absolute magnitudes in the gap
between classical novae and supernovae.  These transients have a set of
observational characteristics which distinguish them from other classes of
optical transients.  These properties are summarized well by
\cite[][hereafter K12]{kasliwal12}: \defcitealias{kasliwal12}{K12}
\begin{itemize}
\item Similar to type I supernovae, showing no hydrogen lines in their spectra
\item Absolute magnitudes in the range from $-15.5$ to $-16.5$
(a factor of about 10~--~30 fainter than the type~Ia supernovae, which
have $M_R$ of about $-19.3$)
\item Characteristic velocities of approximately 6000~--~11000~km~s$^{-1}$
\item Very large calcium abundances, as inferred from their nebular spectra
\item Faster evolution than average type Ia supernovae with a rise time
$\leq 15$~days, whereas type~Ia have rise times $\sim 18$~days
\citep[though it is within the full range of type~Ia rise times found; e.g.][]{hayden10}
\item Current, small sample favors the outskirts of known galaxies
\end{itemize}
As such, many of the properties of these objects are intermediate
between those of classical novae and type~Ia supernovae.  

Previous studies of calcium-rich gap transients have used the common relations for
type~Ia supernovae to estimate some basic parameters of the explosion.
\citetalias{kasliwal12} estimate the masses of the ejecta by setting the ejecta
mass proportional to $v t_r^2$ ($v = $photospheric velocity, $t_r = $rise time),
as first suggested by \cite{arnett82}, and
scaling to type~Ia supernovae.  This yields typical total masses of
$\sim0.3$~--~0.7~M$_\odot$, where the fraction of $^{56}$Ni produced is very
small (e.g. 0.016~M$_\odot$ for PTF~10iuv).  However, \citeauthor{arnett82}'s
modelling explicity requires that there be substantial $^{56}$Ni in the
supernova ejecta, and modelling of the nebular spectrum of SN~2005E in
\citetalias{kasliwal12} gives a mass about half as large as the
\citeauthor{arnett82} formula does.  Type~Ia supernovae are predominantly
powered by radioactive decay of iron-peak elements, where the amount of nickel
produced would be large and the amount of calcium produced would be small
\citep[e.g.][]{woosley86,bildsten07}.  Therefore, the modelling falls short in
properly describing the explosion.

Various models exist to explain the unusual characteristics of these transients,
many of which are summarized in \citetalias{kasliwal12}.  Many of them have at
least one serious flaw that makes them unable to explain these transients.  To
explain the calcium-rich gap transient, SN~2005E, \cite{perets10} concluded that
they were witnessing helium detonation in an interacting double white dwarf (WD)
system with a helium WD mass donor \citep{shen09,waldman11}, but the light curve
is not well-matched.  Various channels for the
detonation of sub-Chandrasekhar-mass WDs exist \citep{woosley94,woosley11}, but
such models do not provide a good match to the observed light curves.
Accretion-induced collapse (AIC) of a rapidly rotating white dwarf into a
neutron star \citep{metzger09} is also ruled out based on the predicted shape of
the light curves, as well as the large velocities and abundances of intermediate
elements predicted.  Recently, \citet{metzger12} presented a one dimensional
nuclear-dominated thick accretion flow (nuDAF) model resulting from the unstable
mass transfer following Roche lobe overflow, but it does not produce the right
amount of helium and calcium to match many of these transients.  Some of the
difficulty in finding a ``best-fit" model may arise in the heterogeneous nature
of the current sample:  there may be selection bias in finding these transients
far from their host galaxies, a range of velocity widths in the lines in the
nebular spectra, different light curve evolutions, for examples.

In \citet{metzger12}, another scenario is mentioned:  the tidal disruption of a
WD by a neutron star or black hole (BH) leading to a nuclear runaway
(detonation).   However, this scenario is not discussed further.  Some other
work has considered the tidal disruption of a WD in other contexts but without
connection to calcium-rich gap transients. \citet{clausen11} considered the
emission spectrum after the disruption but not detonation of a WD by a BH; the
spectrum is incomplete because they did not consider the fusion of heavy
elements in the tidal debris (considered only mass fractions of 67\% oxygen,
32\% carbon, and 1\% helium).  While investigating the vertical compression of
tidally disrupted debris, \citet{stone13} noted the possible importance of WD
tidal disruptions to gravitational wave signals.  Finally, \citet{shcherbakov13}
proposed that a highly unusual combination of a gamma ray burst and a supernova
explosion could be explained by a WD tidal disruption.

In this work we consider the tidal disruption leading to nuclear runaway
(detonation) of WDs in more detail to explain the calcium-rich gap transients.
The goal of the paper is mainly to discuss the {\em plausibility} of
this scenario in connection to calcium-rich gap transients, something that has
not yet been done in the literature.  Future more detailed analyses of this
scenario in the context of Ca-rich gap transients will need to be undertaken
when better multiwavelength (especially high-energy) observations become
available.  These are needed to solve some of the large uncertainties and
degeneracies in various model parameters discussed throughout the text.

In section~\ref{section:origin}, we describe the model and how it can explain
various observational characteristics:  masses, abundances, light curves, and
the positions of these sources.  Then in section~\ref{section:scenarios}, we
discuss some possible scenarios that could produce the tidal detonation and
calculate expected rates of these transients to verify if the proposed origin is plausible.  In
section~\ref{section:SN2012hn}, we then analyze a {\em Chandra} observation of
one of these transients, which enables us to place upper limits on the masses of
the WD and BH in the progenitor system.  Finally in
section~\ref{section:summary}, we summarize our results and suggest
observational tests of our model.

\section{Matching Theory with Observation}
\label{section:origin}

We propose that the model developed by \citet[][hereafter R09]{rosswog09}
\defcitealias{rosswog09}{R09} naturally explains the set of calcium-rich gap
transients.  R09 use a three-dimensional smoothed particle hydrodynamic (SPH)
method to simulate the close interaction between a WD and a more compact object
(a BH) that completely disrupts the WD.  When a WD passes too close to a BH
(within the Roche or tidal disruption radius), the differential gravitational
force from the BH will overcome the gravitational force of the WD.  This tidal
force can elongate the WD in the orbital direction and compress the WD in the
direction perpendicular to the orbit to a high-enough density and pressure to
initiate runaway nuclear burning.  If the nuclear energy is greater than the
binding energy of the star, an explosive nuclear detonation occurs, producing a
luminous transient event.  

The tidal disruption leading to the possible detonation of a WD will only occur
when $M_{BH} \la 2 \times 10^5$~M$_\odot$ or stellar- or intermediate-mass
BHs\footnote{WDs are not tidally disrupted by supermassive BHs because the tidal
disruption radius is inside the event horizon; they are swallowed whole.},
(IMBHs; see R09 for a more detailed discussion).  R09 consider
100~M$_\odot < M_{BH} < 10000$~M$_\odot$, but BHs at a larger range of masses
(up to $M_{BH} \approx 10^5$~M$_\odot$ or $M_{BH} < 100$~M$_\odot$) or even
neutron stars may be viable detonators as well.  Next, we discuss how each of a
set of testable theoretical predictions of this model can be explained by
current observations of this group of transients.

\subsection{The mass of the detonated white dwarf and abundances of leftover material}
\label{section:abundance_mass}

There already exist a set of strong theoretical and observational ways to
constrain the mass of the detonated WD in this scenario.  We first
explore the upper limit to the mass of the WD, which is most strongly motivated
by the abundance patterns in the nebular spectra.

As implied by the name, these calcium-rich gap transients produce a relatively
large amount of calcium.
However, they do not produce a large amount of iron-peak elements (see the spectra
in \citetalias{kasliwal12}).  This is only well-matched to the tidal detonation
scenario if $M_{WD} \la 0.6$~M$_\odot$ \citepalias{rosswog09}.  As
\citetalias{rosswog09} present the most realistic description of a tidal
detonation of a WD to date, this is currently the best estimate of reaction
rates in this scenario.  However, there are some important limitations to this
work.  The upper limit to the WD mass is not very precise because the WD mass
grids presented in \citetalias{rosswog09} are very coarse and might be a
function of other parameters (e.g. tidal radius relative to pericenter,
BH mass).  In addition, the exact boundary of the upper limit is
also not well-determined because the nucleosynthesis calculations presented in
\citetalias{rosswog09} are based on a simplified nuclear reaction network.

This latter point is particularly important because this implies one cannot make
specific abundance predictions from this model.  For instance, silicon is taken
to represent all the elements within the quasi-equilibrium group near silicon in
this network.  Therefore, given that calcium lies within this group, as does
scandium, it should not be a surprise that large amounts of calcium are not
specifically predicted.  The tentative detection of strontium in one
calcium-rich transient \citep{sullivan11} would potentially present a problem,
as strontium cannot be produced in this scenario, but it should be noted that
this detection is of marginal significance and only has been seen in one object.
A clear prediction of this picture is that, as a sample of events
with well-measured abundances emerges, progressively higher ratios of the
iron-peak elements to the elements near silicon in atomic number will be seen as
the peak luminosities rise.  However, since we expect a range of impact
parameters for the tidal detonation model, future work may require a very large
sample size to discern these trends.

In addition to the abundance analysis, which requires that $M_{WD} < 0.6$~M$_\odot$,
the density-dependent Roche limit may favor the tidal disruption of lower-mass WDs.
Lower-mass WDs have lower densities ($\rho$) because they are larger.  Because the
minimum distance at which an object is disrupted by tidal forces scales as $\rho^{-1/3}$,
this may imply that lower-mass WDs are more susceptible to stretching and compression,
possibly making them easier to detonate.  However, based on the limited number of grids
in \citetalias{rosswog09}, it is not clear whether this susceptibility outweighs other
effects (e.g. needing to reach a critical density to detonate, which would require more
compression); more work on this topic is clearly needed.

If this is true, then this would imply that the lowest-mass WDs are be preferred in
this scenario.  This is a problem for single star evolution, as it only produces WDs
with $M \ga 0.45$~M$_\odot$ for the lowest-mass evolved stars in the universe
\citep[e.g.][]{marsh95,kepler07}.  However, binary processes can produce lower-mass
systems \citep[common envelope evolution and Roche lobe overflow; e.g.][]{nelemans01}
and account for the low-mass WDs found in the last couple decades
\citep[e.g.][]{bergeron92,bragaglia95}.  The lowest-mass WDs yet found are
commonly referred to as extremely low mass (ELM) WDs $\sim 0.2$~M$_\odot$ and
are helium-dominated \citep[e.g.][]{liebert04,kilic07,vennes11,hermes13}. Putting
these constraints together, we conclude that
$0.2$~M$_\odot \la M_{WD} \la 0.6 $M$_\odot$.

\subsection{The continuing power source and light curve}
\label{section:light_curve}

Absorption lines in type~Ia SNe tend to become narrower with time, as the outer
layers become more diluted by the expansion of the ejecta and the line-forming
region recedes to lower velocities \citep[e.g.][]{filippenko97,foley11}.  However,
\cite{valenti14} observe the opposite trend in one of the calcium-rich gap
transients, SN~2012hn: constant photospheric velocity (as inferred from the
blueshift of absorption lines) and increasing linewidths, which are better
explained by a continuing energy source.

In supernovae, the radiative decay of nickel serves as a continuing source of
energy after the original ejection of material in the explosion.  However, as
discussed in the previous section, iron-peak elements are not appreciably seen
for calcium-rich gap transients nor would we expect them in this scenario.
\citetalias{rosswog09} find that, in the low-mass WDs ($\sim 0.2$~M$_\odot$),
only about 0.03~M$_\odot$ of iron-peak elements are likely to be produced in the
explosion, and the detonated 0.6~M$_\odot$ model produces an abundance of
iron-peak elements $\sim 4 \times$ less than the detonated 1.2~M$_\odot$ models.

Instead, the power source can be naturally explained by fall-back
accretion of a fraction of the WD material onto the BH.  The very hot inflow of
gas (a few tens to hundreds of million K) not only produces an optical flare but
also produces copius X-ray emission.  The highly blueshifted and broadened lines
(up to thousands of km~s$^{-1}$) can be explained by motions of material associated
with the accretion flow, as seen for both stellar and supermassive
BHs \citep[e.g. disk winds;][]{murray95,miller08,king12,tombesi15}.

For instance, \citetalias{rosswog09} show that for the detonation of a
0.2~M$_\odot$ WD by a 1000~M$_\odot$ BH, the accretion rate onto the central BH
will initially be very large ($\sim 10^{-5.4}$~M$_\odot$~s$^{-1}$) for the first
approximately 10~min and then decay as $t^{-5/3}$ \citep{rees88,phinney89}.
Detailed modelling of the optical light curve using type~Ia templates has been
challenging for all previous work and is beyond the scope of this paper.
Nevertheless, a comparison of the slope of late-time R-band light curve
($\sim 10$~--~100 days after the peak) of a few of the calcium-rich gap transients
with well-sampled light curves (e.g. SN 2005E, PTF~10iuv) indicates that the tidal
detonation scenario ($\dot{M} \propto t^{-5/3}$) seems to provide a better match than
the exponential radioactive decay of $^{56}$Ni as done in \citetalias{kasliwal12},
even when taking into account more of the decay chain
\citep[$^{56}$Co and the kinetic energy loss from its $\beta+$ decay; e.g.][]{dado13}.
Another more more complicated possibility is that the optical light curves could
be explained by the decay of lower-mass elements ($^{48}$Cr $\rightarrow ^{48}$V
$\rightarrow ^{48}$Ti and $^{44}$Ti $\rightarrow ^{44}$Sc
$\rightarrow ^{44}$Ca), producing late-time emission \citep{valenti14}. 

Later on, when the accretion rate falls below the Eddington rate onto this BH,
$L_{Edd} \approx 10^{41} (M_{BH} / 10^3$~M$_\odot$)~erg~s$^{-1}$ or
$\dot{M}_{Edd} = 2.2 \times 10^{-5} (M_{BH} / 10^3$~M$_\odot)$~M$_\odot$~yr$^{-1}$
\citep[standard radiative efficiency 0.1 assumed; e.g.][]{novikov73}, which
occurs approximately seven months after the detonation of a 0.2~M$_\odot$ WD
around a 1000~M$_\odot$ BH in \citetalias{rosswog09}, the X-ray luminosity
should begin to decay.  However, the exact brightness and evolution of the X-ray
bright phase depends a number of poorly constrained parameters:  the WD's
initial mass and trajectory, the mass of the BH, and the structure of the
initially super-Eddington accretion flow \citep[e.g.][]{lodato11,jiang14}, for
examples.

As an illustration of the importance of the assumed structure of the accretion
flow, the viscous accretion timescale is proportional to the ratio of the disk
scale height to tidal disruption radius squared \citep{li02,metzger12}.
Accretion disks can radically vary in scale height depending on the accretion
mode, which is tied to the efficiency of radiative cooling \citep{hayasaki15}.
These uncertainties make the viscous dissipation timescale uncertain by many
orders of magnitude.  This strongly affects our ability to apply upper limits to
physically meaningful parameters ($M_{WD}$, $M_{BH}$) in our X-ray observations
of one of the calcium-rich gap transients in section~\ref{section:SN2012hn}.

\subsection{Positional information}
\label{section:position}

To date, the calcium-rich gap transients have been located predominantly on the
outskirts of their host galaxies (if this is not a selection bias, which is not
clear at the present time).  \citet{yuan13} find that the transients are usually
well-beyond the extent of the nearby host galaxies ($\sim 5 \times$ the K-band
half-light radius on average).  They conclude that these transients do not trace
the stellar mass profile but could be matched to globular clusters or dwarf
galaxies.

Initial non-detection limits for many of the sources ($M_R \ga -10$~--~12~mag)
required that host stellar clusters have $L \la 10^7$~L$_\odot$
\citep{kasliwal12}.  However, recent, much deeper observations searching for
hosts for a couple of the transients have ruled out all but the faintest dwarf
galaxies and globular clusters \citep[$M_R \ga -5.3$~mag for SN~2005E, $M_R \ga -5.6$~mag for SN~2012hn, and $M_R \ga -7.3$~mag for SN~2003H;][]{lyman14}\footnote{SN~2003H's upper limit is not as deep because it lies in a
region of higher background between two merging galaxies.}

For these systems, these observations make the scenario of the tidal detonation
of a WD caused by a random, flyby encounter with an IMBH very
unlikely because the encounter rate would be too small without the presence of
a dense stellar system.  In these cases, interactions producing high-velocity runaway
systems must be considered \citep{foley15}, where the detonations occur in isolated
three-body systems.  We provide a detailed discussion of random encounter rates and
possible three-body interaction scenarios that can satisfy these requirements in
section~\ref{section:scenarios}.

\subsubsection{Can these stellar systems survive?}

To discern whether globular clusters and dwarf galaxies will survive long enough
to be observable in the local universe, we can apply the Roche or tidal
disruption limit to these possible host stellar systems.  The dynamical
friction timescale for galactic cannibalism is \citep{binney08}
\begin{equation}
t_{\rm fric} = \frac{2.7 \ {\rm Gyr}}{{\rm ln} \ \Lambda} \frac{r_i}{30 {\rm \ kpc}} \left( \frac{\sigma_M}{200 \ {\rm km \ s^{-1}}}\right)^2 \left( \frac{100 \ {\rm km \ s^{-1}}}{\sigma_s}\right)^3,
\end{equation}
where $\Lambda$ is the Coulomb logarithm, $r_i$ is the initial radius,
$\sigma_M$ is the typical velocity of these hosts through the halo, and
$\sigma_s$ is the velocity dispersion within a globular cluster or dwarf galaxy.
If the \cite{faber76} law is satisfied, then the value of $\Lambda$ will be
$2^{3/2} \sigma_M/\sigma_s$ \citep{binney08}\footnote{The fainter
end of the dwarf galaxy luminosity function shows some evidence for
increased dark matter fractions and mild deviations from the
fundamental plane for elliptical galaxies \citep[e.g.][]{zaritsky06},
but this should not affect our qualitative conclusions.}.  We pick an
intermediate velocity dispersion between the largest dwarfs and globular
clusters, $\sigma_s = 40$~km~s$^{-1}$.  Using the typical distance of these
objects from their host galaxy ($r_i \sim 25$~kpc, \citetalias{kasliwal12}),
we find that it would take about a Hubble time for these hosts to inspiral if
they start on a circular orbit (and eccentric orbits would only have semi-major
axes larger than this for the objects to have dynamical friction timescales
shorter than a Hubble time).  Therefore, many of these objects will exist in the
semi-local universe.  Because globular clusters typically have higher central
stellar concentrations, they will survive to much smaller galactocentric radii
than dwarf galaxies \citep{lucio06}; however, ultra-compact dwarf galaxies
might be an exception \citep[e.g.][]{strader13}.

\section{Specific Orbital Scenarios and Rates Discussed}
\label{section:scenarios}

As discussed in the previous section, likely host systems for these transients
are faint dwarf galaxies or globular clusters given our scenario.  Precise rate
predictions for this scenario cannot be calculated at the present time due to
considerable uncertainties in multiple key parameters:  the fraction of globular
clusters and faint dwarf galaxies which contain intermediate- or stellar-mass BHs,
the masses of those BHs,
the masses of the WDs that might be detonated, the structural parameters (and
hence central stellar densities) of the clusters and/or faint dwarf galaxies which
contain those BHs, etc.  We only consider rough estimates to test if this model
is plausible or not.  A thorough analysis is well beyond the scope of this paper
and is very challenging because it would have to be quite complex, taking into
account (not an exhaustive list): the binary fraction around the primary BH,
general relativistic corrections, stellar densities, the replenishment rate of
the loss cone of the BH, the spin of the BH, the mass ratio of the BH in the
case of a binary, etc., while having good observational constraints for the
relevant parameters.

We consider two basic plausible scenarios in which a WD could get close enough
to a BH with $M_{BH} \la 10^5$~M$_\odot$ to be tidally disrupted:  1)
the chance flyby of a single WD to an IMBH and 2) the perturbation of a
three-body system, which is especially needed for the stellar-mass BHs with
lower interaction cross-sections ($\propto R_{BH}^2 \propto M_{BH}^2$).  Then we
consider how these ideas fit with the locations of these events.

\subsection{Chance white dwarf flyby of an intermediate-mass black hole}
\label{section:flyby}

Given the larger interaction cross section of an IMBH, chance flybys in dense
stellar systems could occur often enough to be observable calcium-rich gap transients.
Rate estimates testing for plausibility are easier to make for globular
clusters, so we consider them first, despite the fact that the positions of the
transients far from the centers of bright nearby galaxies may favor faint dwarf
galaxies as the sites of the explosions.

\subsubsection{Rates in globular clusters}
\label{section:cluster_rate}

\cite{baumgardt04} estimates a rate of $\sim 10^{-9}$ WD tidal disruptions per
globular cluster per year, assuming typical clusters have a 1000 M$_\odot$ BH
and a central stellar density of $10^5$ stars parsec$^{-3}$. Given $\sim 400$
globular clusters per typical Milky Way-like galaxy \citep[taking a number
larger than that for the Milky Way, since its specific frequency of globular
clusters is relatively low,][]{galleti07}, and a density of galaxies of about
1~Mpc$^{-3}$, then within 200~Mpc, there should be $\sim 10^{10}$ globular
clusters within the distance range out to which the calcium-rich gap transients have
been seen.  If $\sim10\%$ contain IMBHs, then it would be reasonable to expect
$\sim 1$ tidal detonation per year.  Given that the tidal disruption rate goes
as the square of the BH mass \citep{baumgardt04}, the rate would be expected to
increase in the event that there is a broad distribution of BH masses.  The
crude rate calculation can be taken as evidence of plausibility, under the
assumption that some non-negligible fraction of globular clusters contain IMBHs.

At the present time, the evidence that globular clusters contain IMBHs is weak. 
Some claims have been made on the basis of stellar dynamical evidence for an
increase in the charateristic velocities of stars in the centers of a few
clusters \citep[e.g.][]{newell76,gerssen02,gebhardt02,noyola08}.  On the other
hand, in most cases, the data can be equally well explained by mass segregation
of neutron stars and WDs down to the central regions of the cluster, causing an
increase in the mass-to-light ratio without an IMBH
\citep[e.g.][]{illingworth77,baumgardt03}.  Alternatives to radial velocity
searches, such as proper motion searches for dynamical evidence
\citep{mclaughlin06,anderson10} and searches for radio emission
\citep{maccarone04,maccarone05a,derijcke06,bash08,maccarone08,cseh10,lu11,
strader12} have failed to find any IMBHs in Galactic globular clusters.  The M31
cluster G1 does show both dynamical evidence for a BH and radio emission
\citep{ulvestad07}, although more recent radio follow-ups have failed to detect
a source at that position, indicating either that the detection was spurious or
the source is strongly variable \citep{miller-jones12b}.

Despite the lack of evidence thus far for IMBHs in globular clusters, IMBHs may
have been more easily formed through the formation of more massive stars in
lower metallicity environments \citep[e.g.][]{belczynski10}.  This may also be
helpful because it does not require the merger of stellar-mass BHs
\citep[e.g.][]{miller02}, where the gravitational radiation rocket effect
\citep{redmount89} can eject the IMBH from the cluster
\citep{holley-bockelmann08}.  Indeed, deep observations of globular clusters
around the Galaxy and other galaxies over the last few decades have discovered
globular clusters at large galactocentric radii and shown that they have low
metallicities \citep[e.g.][]{searle78,rhode04}.  However, whether or not the
globular cluster radial composition gradient \citep{zinn85} extends
farther out into the halo is not clear \citep{vandenbergh03}.

As discussed Section~\ref{section:abundance_mass}, the tidally disrupted object
must be a low-mass WD and is likely a helium WD.  Unfortunately, the number density
of helium WDs in globular clusters is not well-known.  Helium WDs cannot form at the
present epoch through normal single-star evolution; they can only form through
binary evolutionary processes \citep[see e.g.][]{hansen03}.  In globular
clusters, this can happen through a direct collision between a red giant and a
compact star.  This mechanism has been proposed for producing the ultracompact
X-ray binaries in globular clusters \citep{verbunt87} and is probably at work in
NGC~6397 \citep{grindlay01,taylor01,hansen03,strickler09}.

Placing observational constraints on the density of helium WDs in nearby globular
clusters is extremely difficult because they are so faint and may not be
possible because they should escape from the centers due to mass segregation
\citep[e.g.][]{fregeau02}.  Only in the presence of an IMBH, which appears to be
absent in these systems, is mass segregation quenched because the interactions
with the IMBH dominate over the interactions with other stars \citep{gill08}.

\subsubsection{Rates in dwarf galaxies}
\label{section:dwarf_rate}

Observations clearly show a dearth of dwarf galaxies within a few tens of kpc
and those found in this region are highly disturbed
\citep[e.g. Saggitarius Dwarf, M32, Willman~1, Ursa Major~II;][]{bekki01,johnston95,willman05,munoz10}.
The ``classical'' dwarf spheroidal galaxies orbiting the Milky Way are all
located at least 25~kpc from the Galactic center \citep[e.g.][]{gallagher94}, as
are the new ultra-faint dwarf galaxies \citep[e.g.][]{belokurov10}.  Streams of
past disruption events of dwarf galaxies by the Milky Way clearly suggest this
is still occurring locally \citep[e.g.][]{johnston96a}.  The preferential
locations of faint dwarf galaxies are in contrast to the distribution of globular
clusters, which includes many objects within a few kpc of the Galactic center.
Therefore, the finding that the few calcium-rich gap transients so far are
located at large galactocentric radii provides early suggestive evidence in
favor of faint dwarf galaxies rather than globular clusters as the host sites (barring
a selection bias).

Thus far, the best case for an IMBH in a small dwarf galaxy is that of HLX-1 near
ESO~243-49 \citep{farrell09}.  This system shows X-ray emission that peaks about
$10^{42}$ ergs/sec, shows state transitions like stellar mass BHs
\citep{godet09}, is spectroscopically confirmed to be a member of the cluster of
galaxies against which it is projected \citep{wiersema10}, and recently has been
found to be in a cluster or faint dwarf galaxy which likely has a young population
\citep{farrell12}.  It has been suggested that the companion star to the possible
IMBH may have been tidally captured, given its rather long orbital period
\citep{lasota11}.

Next, we proceed to test this scenario for plausibility as we did for globular
clusters, although the rate estimate for tidal detonations is quite crude
because the input parameters are so poorly constrained.  We use the tidal
disruption rate relation from \cite{baumgardt04}, where the key, variable
parameters are the IMBH mass ($M_{IMBH}$), the central stellar density ($n_c$),
and the central velocity dispersion ($v_c$).  Given an upper limit on the mass
of the host galaxies of many of these events of $\sim 10^7$~M$_\odot$
\citepalias{kasliwal12} and the BH-bulge mass relation
\citep[e.g.][]{mcconnell13}, dwarf galaxies would have $M_{IMBH} \sim \text{a
few} \times 10^4$~M$_\odot$, roughly consistent with the upper limit derived in
Section~\ref{section:SN2012hn}.  A tidal detonation would occur in dwarf
galaxies with dense central stellar clusters like M60-UCD1 \citep{strader13} but
at least a factor of a few fainter.  Considering that the central surface
density of M60-UCD1 is close to but slightly less than that of the densest
globular clusters, we adopt $n_c \sim 10^4$~stars~pc$^{-3}$.  We adopt
$v_c = 40$~km~s$^{-1}$ as in section~\ref{section:position}, assuming that
$\sigma$ is not considerably different from $v_c$.

Assuming the tidal disruption rate roughly scales as
$M_{IMBH}^2 n_c^{7/5} v_c^{-21/5}$ \citep{baumgardt04}, we find that the dwarf
galaxies with IMBHs have a tidal disruption rate per object about 20 times
higher than globular clusters.  Then, assuming 1/10 globular clusters have
IMBHs, we would need only 1/200 as many dense dwarf galaxies as globular
clusters to produce the observed rate of calcium-rich gap transients.  Surveys
of these galaxies are far from complete but \cite{phillipps01}, for example,
estimate that there are $\sim100$ dwarf galaxies in the central region of the
Fornax cluster in the magnitude range where they would be plausible hosts for
IMBHs, and this region contains considerably less than 20000 globular clusters
\citep[e.g.][]{ostrov98}.  Thus, faint, dense dwarf galaxies appear to be a viable
candidate site for tidal detonations of helium WDs by IMBHs for at least some
of the calcium-rich gap transients.

\subsection{Three-body interactions}
\label{section:3-body}

There are a variety of ways in which three body interactions could perturb a WD
to send it into a close encounter with a BH.  Such interactions are frequently
intrinsically chaotic \citep[e.g.][]{portegieszwart14}, even resulting in
orbital flips, for instance \citep{li14}.  Given the very large body of
literature regarding three-body interactions
\citep[see][for summaries]{valtonen06,musielak14}, we only briefly describe
and discuss a few possible scenarios that would induce the tidal disruption of a
WD.  Again, this summary is meant only as a demonstration of plausibility for
three-body interactions as a possible channel for the WD tidal detonations, not
an exhaustive discussion or a detailed analysis of any particular model.

While the study of the three body problem goes back hundreds to years to as
early as the days of Sir Isaac Newton, considerable development on the problem
has occurred in approximately the last half century \citep[e.g.][]{eggleton95}.
Much work has been focused on the well-known Kozai-Lidov mechanism, where perturbations in the
three-body system transfer angular momentum between orbital modes with different
eccentricity and inclination \citep{kozai62,lidov62}.  Since then, numerous
simulations \citep[e.g.][]{anosova90} have been carried out to explore the large
family of solutions that exist, from larger N-body encounters \citep{leigh12} to
higher order expansions \citep[octupole,][]{naoz13}.  The importance of
three-body interactions has also been explored for a variety of astrophysical
contexts outside of the solar system (e.g. ultra-luminous X-ray sources,
\citealt{blecha06}; blue straggler stars in dense globular clusters,
\citealt{chatterjee13}; gravitational waves, \citealt{antonini14}; the final
parsec problem involving the merger of supermassive BHs, \citealt{khan13}; the
motions of stars around BHs in galaxy centers,
\citealt{sesana06,sesana07,sesana08}; extrasolar planets, \citealt{mardling13};
white dwarf mergers, \citealt{hamers13}).
We now consider its importance for the tidal disruption of WDs when they are
paired with a BH and another member (another star or remnant) in a triple
system.  We consider paths to exchange angular momentum which do and do not
involve mass transfer.

First, we consider a few examples of three-body interactions that will exchange
angular momentum without mass transfer.  The first two options require the
Kozai-Lidov mechanism and are enhanced in higher-mass-ratio systems.  Rapid
eccentricity oscillations \citep{ivanov05} can make the magnitude of angular
momentum in the inner binary comparable to the total angular momentum of the
system.  When numerically evolving the triple system, \cite{antognini14} find
that, because the general relativistic terms are sensitive to the orbital
eccentricity, the inner binary can
merge well within a few Gyr, or even faster if the mass ratio is large because
it is proportional to the amplitude of the eccentricity oscillations; this would
be the case for a WD and BH inner binary. Second, a BH binary system in a
stellar cusp strongly enhances the rate of tidal disruption over two-body
relaxation for a single BH \citep{chen09,chen11}.  This was originally
demonstrated for a supermassive BH and another BH in a binary in the center of a
stellar cusp of a galaxy, but also applies to a stellar-mass BH and IMBH pair in
the center of a globular cluster or dwarf galaxy, which are relevant for calcium-rich
gap transients.  The tidal disruption rate in this case is at least
approximately an order of magnitude larger than a single IMBH fed by two-body
relaxation \citep{wang04} but it also quickly depletes the loss cone in $\la
10^6$~yr.

Yet another way a triple system can cause a WD in the
inner binary to migrate toward a BH is through the tidal excitation of high
frequency p-modes.  These can redistribute angular momentum in a triple system
\citep{fuller13}, causing the inner binary to inspiral.  Finally, the
redistribution of angular momentum does not have to occur within a triple star
system to significantly perturb a close binary.  A close flyby of another
star/remnant or binary can induce a high-enough eccentricity in a WD-BH binary
that forces the WD to pass within the tidal disruption radius
\citep[see][for detailed discussions]{heggie96,maccarone05d}.

Finally, we consider two examples which involve mass transfer.  In the first of
these cases, \cite{stahler10} show that the mass transfer/accretion of gas from
a dense interstellar medium (star-forming region) can act as a way to transfer
angular momentum.  When moving through and accreting the dense gas, the orbit
can be shrunk by dissipating angular momentum through acoustic waves.  This
would require the site of the transient to be near a dense molecular cloud;
however, these transients are not generally associated with star formation
\citep{lyman13}.  Another possibility is that mass is transferred from star to
star within the binary or lost as the stars evolve off of the main sequence
\citep{michaely14}.  This can happen in a much older stellar population and
induce a chaotic orbital behavior.

As has just been made clear, there are a multitude of various ways three-body
interactions can lead to the tidal disruption of a WD, many of which do not
require extremely dense stellar systems \citep[e.g. the importance of three-body
interactions in lower-density stellar groups;][]{leigh13}.  This makes it very
difficult if not impossible to rule out the tidal disruption origin on
interaction cross-section arguments alone.

\section{The particular case of SN 2012hn}
\label{section:SN2012hn}

\begin{figure}
 \includegraphics[width=\columnwidth]{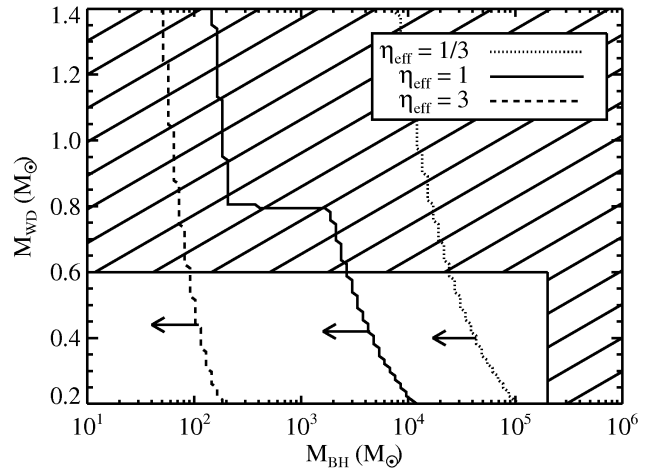}
 \caption{Joint upper limits for the mass of the WD and BH involved in the
 tidal detonation of SN~2012hn.  Our X-ray observation upper limit excludes
 parameter space to the upper right of the curves.  A description of
 $\eta_{eff}$ and how these limits are calculated is provided in
 Section~\ref{section:SN2012hn}.  The horizontal jump that appears on the plot
 for $\eta_{eff} = 1$ is the hard/soft-state transition (the state
 transition is not crossed on the other plotted curves).  The excluded parameter
 space, where the WD is swallowed whole or the WD produces too many heavy elements,
 is discussed in section~\ref{section:SN2012hn} (note:  the boundaries are
 approximate).}
 \label{fig:upper_limits}
\end{figure}

To test the proposed origin of the calcium-rich gap transients as tidal
disruptions of WDs, we observed one of these sources, SN~2012hn
\citep{valenti14} to look for accretion emission during the decay phase.
SN~2012hn has all of the typical characteristics of calcium-rich gap transients
as outlined in Section~\ref{section:intro}.  It is also fairly nearby, at a
distance of only 27~Mpc, in the outskirts of the galaxy NGC~2272.

Since accretion is observed most unambiguously in the X-rays, we used the
{\em Chandra X-ray Observatory}.  We observed SN~2012hn with {\em Chandra} for a
continuous 30~ksec on 17~August~2013 (OBS\_ID=15668), 533 days after the
explosion, which occurred on 12~March~2012.  Data were taken in timed exposure
mode and telemetered in very faint mode.  Data reduction and analysis were
completed using CIAO version 4.6 \citep{fruscione06} and Sherpa version 2
\citep{freeman01}.  We reprocessed the level=1 event file using the
{\sc chandra\_repro} script to apply the most recent calibration updates
available (CALDB version 4.6.1.1).

The source was not detected in the observation.  To calculate upper limits
to the source flux, we extracted source and background spectra using
{\sc specextract} at the location of the transient in a $6^{\prime \prime}$
radius circle and a $25^{\prime \prime}$ radius annulus excluding the source
region, respectively.  We jointly fit the unbinned source and background
spectra at 0.5~--~8.0~keV using the C-statistic, which is similar to the
\cite{cash79} statistic but with an approximate goodness-of-fit measure.
Since we constrain $ L_X \la 0.02 L_{Edd}$ for any BH mass we consider, the
accretion disk is in the canonical low/hard state \citep[e.g.,][]{maccarone03c},
in which the spectrum is usually well-described by a power-law model.  For the
source spectral model ({\sc xsphabs $\times$ xspowerlaw}), we set the photon
index of the power-law, $\Gamma = 1.6$.  This
value is typical for accretion in average X-ray binary populations
\citep[e.g.][]{sell11}.  The column density, $N_H$ was frozen at the galactic
foreground value \citep{dickey90}.  The background spectrum was fit with a
$\Gamma = 1.4$ power law, consistent with the hard X-ray background
\citep[e.g.][]{tozzi06}.  We find a $3 \sigma$ upper limit to the unabsorbed
flux, $F_{0.5-8 keV} < 5.1 \times 10^{-15}$~erg~cm$^2$~s$^{-1}$.  At a distance
of 27~Mpc, this corresponds to $L_{0.5-8 keV} < 4.4 \times 10^{38}$~erg~s$^{-1}$.

To put this upper limit in context, we adopt the analytical relations from
\citet{li02} for the similar case of the tidal disruption of a star by a
supermassive BH to calculate joint upper limits on the mass of the WD and BH
assumed to be involved in the disruption (given the X-ray luminosity upper limit
above at the time of the observation).  While more complex numerical models
exist \citep[e.g.][]{piran15}, we opt to use the simpler analytical formulation
to derive parameter limits, especially since the source is not X-ray-detected.

First, we combine equations (10) and
(11) from \citet{li02} to calculate the fallback accretion rate, $\dot{M}$, as a
function of time, $t$, fallback fraction, $f$, and mass of the BH, $M_{BH}$.  We
then calculate the accretion efficiency, $\epsilon$, using equation (12) from
\citet{li02}.  In addition, we make a few more assumptions that simplify or
improve our model:
\begin{itemize}

\item We use the mass-radius relation for WDs, assuming the mean molecular
weight per free electron, $\mu_e = 2$ \citep[e.g.][their equation 7.103]{hansen04}.

\item We assume that $f = 0.35$, as suggested by \citetalias{rosswog09}.

\item We take into account the accretion disk state transition of accreting BH
binaries so that $L \propto \dot{m}$ for $L/L_{Edd} \geq 0.02$ and
$L \propto \dot{m}^2$ for $L/L_{Edd} < 0.02$ \citep[e.g.][]{esin97,maccarone03c}.

\item To account for the fraction of light falling in the {\em Chandra}
bandpass, we assume a bolometric correction, $\delta$, of 0.85 in the high, soft
state and 0.1 in the low, hard state \citep[e.g.][]{portegieszwart04}.

\end{itemize}

To take into account that the various scale factors are jointly uncertain by
approximately a factor of a few, we define a relative effective efficiency:
\begin{equation}
\eta_{eff} = \left( \frac {f} {0.35} \right) \left( \frac {\delta_{soft,hard}} {0.85,0.1} \right) \epsilon
\end{equation}

We plot the joint upper limits in $M_{BH}$ and $M_{WD}$ for a range in a factor
of 3 for $\eta_{eff}$ in Fig.~\ref{fig:upper_limits}.  The effects for
different BH spins, choice of X-ray spectral model and other uncertainties
discussed in section~\ref{section:light_curve} are also subsumed into the
uncertainty in $\eta_{eff}$.  The fact that $0.2 \la M_{WD} \la 0.6$ and
M$_{BH} \la 2 \times 10^5$~M$_\odot$, which excludes large sections of
parameter space given the tidal detonation model, are discussed in
section~\ref{section:abundance_mass} and the beginning of
section~\ref{section:origin}.

From this analysis, we have placed weak constraints on the BH mass.
It is clear that these observations are not deep enough or taken close enough to
the time of the transient to place strongly constraining upper limits on the host
system under this model.

\section{Summary and tests of the model}
\label{section:summary}

In this work, we have proposed that tidal detonations of low-mass WDs are the
source of at least some of the calcium-rich gap transients.  We have shown that
the predictions of this model are at least qualitatively consistent with
observations of the systems discovered thus far.  Throughout this work, we have
assessed this model using rough
comparisons to test for plausibility; however, because of the large range of
ways in which the tidal detonation can arise and the large uncertainties in key
parameters in these systems, detailed quantiative comparisons are not possible
at the present time and currently may not be computationally feasible.  

A few clear, testable predictions can be made if this scenario is correct:

\begin{enumerate}

\item If these transients are random flyby encounters of low-mass WDs with IMBHs,
most of the calcium-rich gap transients should be associated with
low-luminosity, preferably dense stellar groups.  However, if three-body
interactions drive a WD already orbiting a BH into it, then these systems could be
anywhere and could include ejected systems \citep[e.g.][]{foley15}.

\item If very faint dwarf spheriodals or ultra-faint dwarfs are the hosts of the
calcium-rich gap transients and are distributed like the brighter dwarf
ellipticals, then these explosions might occur preferentially in clusters or
dense groups of galaxies \citep{vader91,sabatini03}.  Of the three most recent
transients presented by \citetalias{kasliwal12}, two were located in
clusters of galaxies, as was one of the two archival events.

\item X-ray emission should be seen from the gap transients, and should follow a
$t^{-5/3}$ decay law.

\end{enumerate}

Using this latter prediction, we sought to catch one of these calcium-rich gap
transients, SN~2012hn, in the decay phase of the light curve.  Unfortunately, we did not
detect the source in our follow-up X-ray observations.  However, assuming our
hypothesis is correct that this transient was a tidal detonation event of a WD,
M$_{BH} \la 10^5$~M$_\odot$ and M$_{WD} \la 0.6$~M$_\odot$
in this scenario.  The upper limits are not very constraining because the
effective efficiency that includes the accretion efficiency, fallback
efficiency, and the bolometric correction is highly uncertain.  If this scenario
is correct, a future X-ray observation of one of these sources as close as
possible to the time of the transient might detect it and provide much stronger
constraints on the masses and the nature of the event.

Finally, concerning the first two points above, some initial work has been
completed to search for the hosts of a few of these transients without success
\citep{lyman14}.  These recent observations rule out for a few sources, including SN~2012hn, that the
transient was caused by the tidal detonation induced by a random flyby of an
IMBH because the upper limits to the masses of host stellar associations are too
low.  However, the current sample of calcium-rich gap transients might be
heteorgeneous, and \citeauthor{lyman14} only placed stringent upper limits for a
few sources.  In addition, the stringent upper limits rule out most but not all
globular clusters and dwarf galaxies.  Finally, there are numerous ways
three-body interactions could perturb the orbits of a stellar system to force a
WD too close to a stellar-mass BH or neutron star.  Future detailed,
multiwavelength observations are clearly needed to constrain this and other
various models for these transients.

\section{Acknowledgments}

We are grateful to Giuseppe Lodato, Cole Miller, Hagai Perets, and Denija
Crnojevic for useful discussions.  This work has been funded through
{\em Chandra} grant GO4-15061X.

\bsp

\label{lastpage}

\end{document}